\documentclass[aps,prl,twocolumn,nofootinbib,showpacs,superscriptaddress,longbibliography]{revtex4-2}

\usepackage{graphicx}
\usepackage{color}
\usepackage{epstopdf}

\usepackage{amsmath,amssymb,bm}
\usepackage{mathrsfs}
\usepackage{dcolumn}

\usepackage{booktabs}
\usepackage{float}
\usepackage{subfigure}
\usepackage{soul}
\usepackage[switch]{lineno}

\usepackage[colorlinks=true,linkcolor=blue,citecolor=blue]{hyperref}
\DeclareGraphicsExtensions{.pdf, .png, .jpg, .eps}

\begin{document}

\title{Cascade of Spin Liquids in a Bilayer Triangular-lattice Antiferromagnet Rb$_2$Co$_2$(SeO$_3$)$_3$}

\author{Xiaoyu~Xu}
\thanks{These authors contributed equally to this study.}
\affiliation{School of Physics and Beijing Key Laboratory of  Opto-electronic Functional Materials $\&$ Micro-nano Devices, Renmin University of China, Beijing, 100872, China}

\author{Yunlong~Wang}
\thanks{These authors contributed equally to this study.}
\affiliation{School of Physics and Beijing Key Laboratory of  Opto-electronic Functional Materials $\&$ Micro-nano Devices, Renmin University of China, Beijing, 100872, China}

\author{Xuejuan~Gui}
\thanks{These authors contributed equally to this study.}
\affiliation{School of Physics and Beijing Key Laboratory of  Opto-electronic Functional Materials $\&$ Micro-nano Devices, Renmin University of China, Beijing, 100872, China}

\author{Jun Luo}
\thanks{These authors contributed equally to this study.}
\affiliation{Institute of Physics, Chinese Academy of Sciences, and Beijing National Laboratory for Condensed Matter Physics, 100190, Beijing, China}

\author{Guijing~Duan}
%\thanks{These authors contributed equally to this study.}
\affiliation{School of Physics and Beijing Key Laboratory of  Opto-electronic Functional Materials $\&$ Micro-nano Devices, Renmin University of China, Beijing, 100872, China}

\author{Ke Shi}
\affiliation{Anhui Key Laboratory of Low-Energy Quantum Materials and Devices, High Magnetic Field Laboratory, HFIPS, Chinese Academy of Sciences, Hefei, Anhui 230031, China}

\author{Zhaosheng Wang}
\affiliation{Anhui Key Laboratory of Low-Energy Quantum Materials and Devices, High Magnetic Field Laboratory, HFIPS, Chinese Academy of Sciences, Hefei, Anhui 230031, China}

\author{Shuo Li}
\affiliation{Institute of Physics, Chinese Academy of Sciences, and Beijing National Laboratory for Condensed Matter Physics, 100190, Beijing, China}

\author{Huifen Ren}
\affiliation{Institute of Physics, Chinese Academy of Sciences, and Beijing National Laboratory for Condensed Matter Physics, 100190, Beijing, China}

\author{Chuanying Xi}
\affiliation{Anhui Key Laboratory of Low-Energy Quantum Materials and Devices, High Magnetic Field Laboratory, HFIPS, Chinese Academy of Sciences, Hefei, Anhui 230031, China}

\author{Langsheng Ling}
\affiliation{Anhui Key Laboratory of Low-Energy Quantum Materials and Devices, High Magnetic Field Laboratory, HFIPS, Chinese Academy of Sciences, Hefei, Anhui 230031, China}

\author{Zhanlong Wu}
\affiliation{School of Physics and Beijing Key Laboratory of  Opto-electronic Functional Materials $\&$ Micro-nano Devices, Renmin University of China, Beijing, 100872, China}

\author{Ying Chen}
\affiliation{School of Physics and Beijing Key Laboratory of  Opto-electronic Functional Materials $\&$ Micro-nano Devices, Renmin University of China, Beijing, 100872, China}

\author{Xiaohui Bo}
\affiliation{School of Physics and Beijing Key Laboratory of  Opto-electronic Functional Materials $\&$ Micro-nano Devices, Renmin University of China, Beijing, 100872, China}

\author{Xinyu Shi}
\affiliation{School of Physics and Beijing Key Laboratory of  Opto-electronic Functional Materials $\&$ Micro-nano Devices, Renmin University of China, Beijing, 100872, China}

\author{Kefan Du}
\affiliation{School of Physics and Beijing Key Laboratory of  Opto-electronic Functional Materials $\&$ Micro-nano Devices, Renmin University of China, Beijing, 100872, China}

\author{Rui Bian}
\affiliation{School of Physics and Beijing Key Laboratory of  Opto-electronic Functional Materials $\&$ Micro-nano Devices, Renmin University of China, Beijing, 100872, China}

\author{Jie Yang}
\affiliation{Institute of Physics, Chinese Academy of Sciences, and Beijing National Laboratory for Condensed Matter Physics, 100190, Beijing, China}

\author{Yi Cui}
\email{cuiyi@ruc.edu.cn}
\affiliation{School of Physics and Beijing Key Laboratory of  Opto-electronic Functional Materials $\&$ Micro-nano Devices, Renmin University of China, Beijing, 100872, China}
\affiliation{Key Laboratory of Quantum State Construction and Manipulation (Ministry of Education), Renmin University of China, Beijing, 100872, China}

\author{Rui Zhou}
\email{rzhou@iphy.ac.cn}
\affiliation{Institute of Physics, Chinese Academy of Sciences, and Beijing National Laboratory for Condensed Matter Physics, 100190, Beijing, China}

\author{Jinchen Wang}
\email{jcwang\_phys@ruc.edu.cn}
\affiliation{School of Physics and Beijing Key Laboratory of  Opto-electronic Functional Materials $\&$ Micro-nano Devices, Renmin University of China, Beijing, 100872, China}
\affiliation{Key Laboratory of Quantum State Construction and Manipulation (Ministry of Education), Renmin University of China, Beijing, 100872, China}
\affiliation{PSI Center for Neutron and Muon Sciences, 5232 Villigen PSI, Switzerland}
\affiliation{Laboratory for Quantum Magnetism, Institute of Physics, École Polytechnique Fédérale de Lausanne (EPFL), 1015 Lausanne, Switzerland}

\author{Rong Yu}
\email{rong.yu@ruc.edu.cn}
\affiliation{School of Physics and Beijing Key Laboratory of  Opto-electronic Functional Materials $\&$ Micro-nano Devices, Renmin University of China, Beijing, 100872, China}
\affiliation{Key Laboratory of Quantum State Construction and Manipulation (Ministry of Education), Renmin University of China, Beijing, 100872, China}

\author{Weiqiang~Yu}
\email{wqyu\_phy@ruc.edu.cn}
\affiliation{School of Physics and Beijing Key Laboratory of  Opto-electronic Functional Materials $\&$ Micro-nano Devices, Renmin University of China, Beijing, 100872, China}
\affiliation{Key Laboratory of Quantum State Construction and Manipulation (Ministry of Education), Renmin University of China, Beijing, 100872, China}

\date{\today}

\maketitle

{\bf In frustrated Ising magnets, classical spin liquids (CSLs) with macroscopic ground-state degeneracy can survive against conventional magnetic order, as exemplified by systems on triangular, kagome and pyrochlore lattices at zero field. Here we report the discovery of a high-field route toward spin liquids in a bilayer triangular lattice antiferromagnet, Rb$_2$Co$_2$(SeO$_3$)$_3$. We demonstrate that a cascade of CSLs$\text{\textemdash}$characterized by doubly degenerate one-up-one-down local spin configurations and a residual entropy of $1/2(1-M/M_{\rm s})R{\ln}2$ per mole$\text{\textemdash}$emerges through field-controlled dilution of Ising dimers. Owing to the interplay of intra- and inter-layer interactions, these CSLs are further stabilized by lattice symmetry breaking at fractional magnetization plateaus. Such field-induced spin liquids can be understood as a consequence of generalized ice rules, analogous to those governing in pyrochlore antiferromagnets. In particular, the 5/6-plateau state is a candidate quantum spin liquid. Our results thereby establish a new pathway for exploring diverse spin liquid states across both classical and quantum regimes.
}

Frustrated magnets offer a fertile ground for realizing unconventional states of matter, as competing exchange interactions can suppress long-range magnetic order and give rise to novel disordered states~\cite{Fazekas_1974,Moessner_PT_2006,kitaev_AP_2006, Balents_Nature_2010,gen_NC-2025}. Among these, classical spin liquids (CSLs) are distinguished by macroscopic ground-state degeneracy and an extensive residual entropy~\cite{Giauque_JACS_1936,Garanin_PRB_1999,wang_PRL_2025}. A canonical example is the Ising triangular lattice antiferromagnet, where, according to the theory~\cite{Wannier_PR_1950}, the two-up-one-down and the one-up-two-down spin-configuration population yields 
a residual entropy $0.338R$ at zero temperature, $R$ being the molar gas constant. Similarly, the Ising kagome antiferromagnet has a residual entropy of $0.502R$~\cite{Kanô_PoTP_1953}. The Ising pyrochlore materials such as Dy$_2$Ti$_2$O$_7$ and Ho$_2$Ti$_2$O$_7$, the magnetic two-in-two-out ice rule governs the ground state in each local tetrahedron, leading to a Pauling-type residual entropy of $1/2R\ln 3/2$~\cite{Harris_PRL_1997,Ramirez_Nature_1999, Siddharthan_PRL_1999,Steven_Sci_2001}. Similar ice-rule behavior has also been observed in the kagome compound HoAgGe~\cite{Zhao_Sci_2020,zhao_ArXiv_2025}. It has been suggested that introducing quantum fluctuations into CSLs may further stabilize quantum spin liquids (QSLs)~\cite{Balents_Nature_2010,Ross_PRX_2011}. Despite considerable theoretical advances in various model systems~\cite{Shannon_PRL_2012,Benton_PRB_2012, Banerjee_PRL_2008,Hermele_PRB_2004, Castelnovo_Nature_2008,Savary_PRL_2012, Yan_Sci_2011,Savary_RPP_2016}, the experimental realization of CSLs and QSLs in real materials remains a significant challenge~\cite{Balents_Nature_2010, Wen_npj_2019,Senthil_sci_2020,xing_CM_2024}. 

Recently, a class of spin-1/2 bilayer triangular-lattice antiferromagnets (TLAFMs), K$_2$Co$_2$(SeO$_3$)$_3$ and Rb$_2$Co$_2$(SeO$_3$)$_3$, was discovered to exhibit strong easy-axis anisotropy~\cite{Zhong_2020_PRB,Xu_2024_CM,Li_PRB_2025}. While several long-range magnetically ordered states have been proposed for its zero-field ground state~\cite{Chen_PRL_2024, Fu_ArXiv_2025}, the precise nature of this state remains under debate. Under a longitudinal field, K$_2$Co$_2$(SeO$_3$)$_3$ displays successive magnetization plateaus at the fractional values $1/3$, $1/2$, $2/3$, and $5/6$ of the saturation magnetization~\cite{Xu_2024_CM}. Beyond the well-established 1/3-plateau commonly observed in TLAFMs, the microscopic origins of the higher-field plateaus are not yet well understood.

\begin{figure*}[t]
\includegraphics[width = 18 cm]{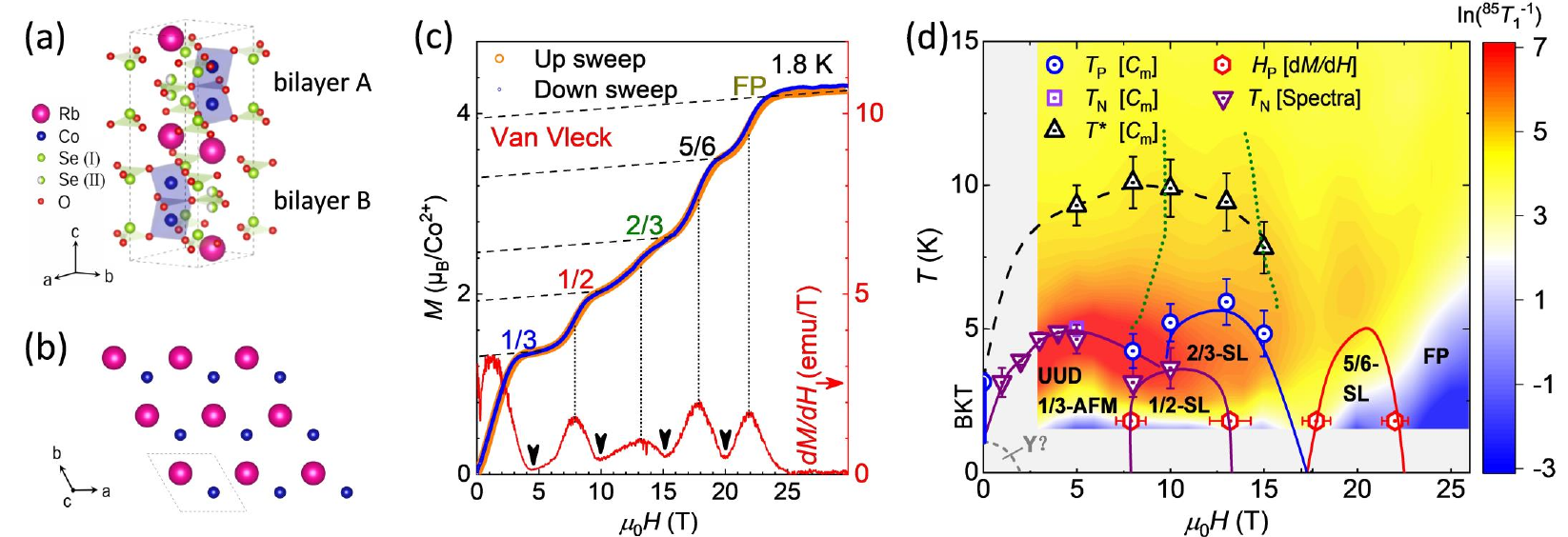}
\caption{\label{pd}{\bf Crystal structure, magnetization plateaus and phase diagram of Rb$_2$Co$_2$(SeO$_3$)$_3$.}
(a) Side view of a unit cell consisting of two Co$_2$O$_9$ bilayers. One selenium atom occupies either of the two adjacent Se(II) sites indicated by half-filled green circles.
(b) Top view of a Co$^{2+}$ bilayer with Rb$^{+}$ ions situated above or below the layers.
(c) Magnetization $M(H)$ measured under a dc field applied along the crystalline $c$ axis. Differential susceptibility $dM/dH$ is derived from upsweep data. Downward arrows mark the dips in $dM/dH$, which define the 1/3-, 1/2-, 2/3- and 5/6-plateau phases. Peaks in $dM/dH$ determine the transition fields between plateau phases. The short-dashed lines alongside the $M(H)$ curves note the Van Vleck contributions.
(d) Phase diagram determined by various probes as indicated, overlaid on a color map of $1/^{85}T_1$  data. $T_{\rm N}$ denotes the AFM ordering temperature. $H_{\rm P}$($T_{\rm P}$) indicates the transition field (temperature) for each plateau phase.  $T^{*}$ marks a crossover to enhanced spin fluctuations upon cooling. The 1/3-, 1/2-, 2/3- and 5/6-plateau phases correspond respectively to the UUD, 1/2-SL, 2/3-SL and 5/6-SL phases illustrated in Fig.~\ref{states}b; their phase boundaries are shown as solid lines. The color contour map of $1/^{85}T_1$ highlights low-energy spin fluctuations. 
}
\end{figure*}

To address these open questions, we performed high-field magnetization, specific heat, and nuclear magnetic resonance (NMR) spectroscopic measurements on Rb$_2$Co$_2$(SeO$_3$)$_3$. Our magnetization measurements confirm distinct magnetization plateaus at 1/3, 1/2, 2/3, and 5/6 of the saturation value. Analysis of the NMR lineshapes unambiguously demonstrates the formation of an up-up-down (UUD) antiferromagnetic order in the 1/3-plateau phase, partial magnetic ordering in the 1/2-plateau phase, and the absence of long-range magnetic order in both the 2/3- and 5/6-plateau phases. The total magnetic entropy $S_{\rm m}$, derived from the magnetic specific heat $C_{\rm m}$ above 1.8~K, exhibits a continuous decrease with increasing field as the system is driven beyond the 1/3-plateau phase. This pronounced entropy reduction observed in the absence of magnetic order is a signature of a spin liquid (SL), arising from a macroscopic ground-state degeneracy enforced by the one-up-one-down rule constraint among the populated interlayer antiferromagnetic Ising dimers. Our theoretical modeling and Monte Carlo simulations confirm the existence of SLs in the 2/3- and 5/6-plateaus, and reveal that the partially ordered 1/2-plateau consists of an alternating pattern of UUD and 2/3-plateau bilayers. These results establish the $S$=1/2 bilayer Ising TLAFM as a novel and compelling platform for exploring CSL physics, as well as potential QSL behavior when quantum fluctuations become significant.
\\

\noindent
{\large \bf Method}\\
\noindent
Single crystals of Rb$_2$Co$_2$(SeO$_3$)$_3$ were grown by the solid-state method following the procedure reported in Ref.~\cite{Zhong_2020_PRB}.
X-ray diffraction measurement on the ground single crystals confirmed that the lattice structure and verified that the flat surface of the as-grown crystals corresponds to the $ab$ plane.

The high-field magnetization were measured at 1.8~K with field up to 35~T, using a vibrating sample magnetometer (VSM) by the water-cooled resistive magnet WM5 at the Steady High Magnetic Field Facility (SHMFF). A Physical Property Measurement System (PPMS) at the Synergetic Extreme Condition User Facility (SECUF) is used to measure the specific heat with field from 0~T to 15~T, and a water-cooled resistive magnet WM2 at SHMFF is used measure the specific heat with field from 5~T to 20~T, respectively.

NMR measurement were performed on $^{85}$Rb nuclei ($I$ = 5/2, Zeeman factor $^{85}\gamma$ = 4.111~MHz/T) and $^{87}$Rb nuclei ($I$ = 3/2, $^{87}\gamma$ = 13.931~MHz/T) with the magnetic field applied along the crystalline $c$ axis. 
For a low-field of 2~T, $^{87}$Rb spectra were collected, taking advantage of its large Zeeman factor to improve the signal to noise ratio. For higher fields, $^{85}$Rb spectra and $1/^{85}T_1$ were collected, taking advantage of its smaller Zeeman factor and therefore narrow NMR bandwidth and slower $T_1$ to gain better data coverage.

For NMR, the sample was cooled using a Variable Temperatures Insert (VTI) with the temperature down to 1.6~K. NMR measurements above 16 T were performed at the SECUF using a 26~T all-superconducting magnet. Spectra were acquired using the standard spin-echo technique. The spin-lattice relaxation time $T_1$ was obtained using the inversion-recovery method. The nuclear spin recovery were fit to the exponential function
$M(t)/M(\infty)$$=\\$1$-$$a[0.03e^{-(t/T_1)^\beta}$$+$$0.18e^{-(6t/T_1)^\beta}$$+$$0.79e^{-(15t/T_1)^\beta}]$, 
where $\beta$ is a stretching factor ($\beta$=1 in the paramagnetic phase).\\

\begin{figure*}[t]
\includegraphics[width = 17 cm]{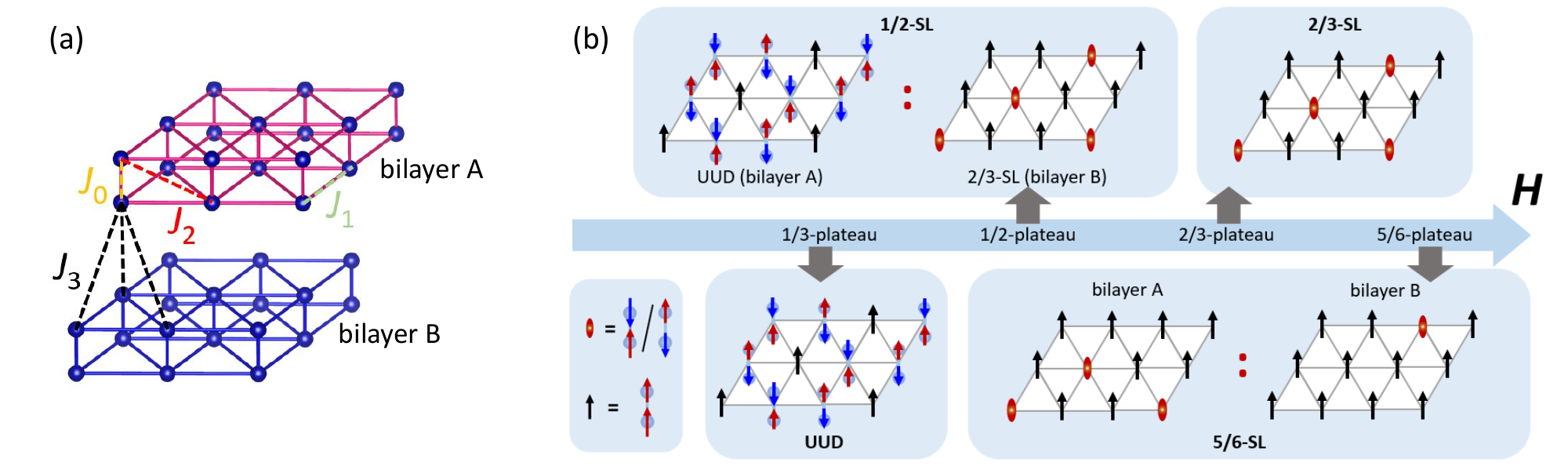}
\caption{\label{states}{\bf Modeling and magnetic states in the plateau phases from Monte Carlo calculations.}
(a) Model of the AB-stacked magnetic bilayers; pink and blue bonds highlight the Co$^{2+}$ exchange networks. All dominant couplings are antiferromagnetic: the Ising intra-dimer coupling $J_0$, intra-layer coupling $J_1$, inter-dimer coupling $J_2$, and inter-bilayer coupling $J_3$.
(b) Illustrations of the magnetic states in each plateau phase. The small panel at the lower left shows two spin configurations on a dimer: the non-polarized dimer with a doubly degenerate one-up-one-down state (represented by a vertical ellipse), and the field-polarized dimer with a two-up-spin state (represented by a upward black arrow). The main panel displays the magnetic configurations at each magnetization plateau under increasing field, calculated using the specific parameters (see Sec.~S1 in the SM~\cite{SM}). The 1/3-plateau exhibits an ordered UUD pattern. The 2/3-plateau hosts a magnetically disordered 2/3-SL, in which polarized dimers form a honeycomb lattice. The 1/2-plateau consists of AB bilayers with alternating UUD and 2/3-SL configurations, referred to as the 1/2-SL. The 5/6-plateau, termed the 5/6-SL, consists of alternating bilayers hosting a 2/3-SL and fully polarized configurations; in addition, degenerate non-polarized dimer can hop among the bilayers.
}
\end{figure*}

\noindent
{\large \bf Results}
\\
\noindent
{\bf Structure, magnetization and phase diagram}
\\
\noindent
Figure~\ref{pd}a illustrates a unit cell of Rb$_2$Co$_2$(SeO$_3$)$_3$, composed of AB-stacked bilayers (labeled A and B) of face-sharing CoO$_6$ octahedra. A top view of the crystal structure (Fig.~\ref{pd}b) highlights a triangular lattice of Co$^{2+}$ ions with an intralayer Co-Co distance $d_1 = 5.52$~\AA. In contrast, Co-Co dimers within each bilayer are aligned along the $c$ axis and are much closer, with $d_0 = 2.93$~\AA. This short intra-dimer spacing suggests a dominant coupling inside the dimer.

We confirmed the strong Ising magnetic anisotropy in the material by the high-temperature susceptibility data (see Fig.~S4 in the SM~\cite{SM}). The magnetization $M(H)$, measured at 1.8~K in dc field up to 30~T, is shown in Fig.~\ref{pd}c. After subtracting the Van Vleck contribution, a fully polarized (FP) phase is established above 25~T. Below this phase, four distinct plateaus are clearly resolved at fractional magnetizations of 1/3, 1/2, 2/3, and 5/6, as indicated. The differential susceptibility, $dM/dH$, provides a precise means to locate these plateaus: dips in $dM/dH$ mark their center fields, while peaks corresponding to the phase boundaries at 1.8~K. 
\\

\noindent
{\bf Theoretical analysis and simulations}
\\
\noindent
Given the strong easy-axis anisotropy (see Fig.~S4 in the SM~\cite{SM}), we model Rb$_2$Co$_2$(SeO$_3$)$_3$ using the following spin-1/2 Hamiltonian (see Sec.~S1 in the SM~\cite{SM}),
\begin{align}
    \label{equ1}
    H &= \sum_i[J_0^{zz}S_{i1}^zS_{i2}^z+\frac{J_0^{xy}}{2}(S^+_{i1}S^-_{i2}+S^-_{i1}S^+_{i2})] \nonumber \\
    &+ \sum_{\langle ij\rangle,\alpha\beta, n}J_n^{zz}S_{i\alpha}^zS_{j\beta}^z - h\sum_{i,\alpha}S_{i\alpha}^z,
\end{align}
where the antiferromagnetic exchange interactions $J_n$ ($n=0,1,2,3$) are illustrated in Fig.~\ref{states}a. The dominant Ising coupling $J_0^{zz}$ favors the doubly degenerate one-up-one-down spin configuration within an interlayer dimer, as shown in the lower panel of Fig.~\ref{states}b, while the magnetic field polarizes the dimer spins into a two-up configuration. The frustrated Ising-type interdimer interactions yield SL ground states under magnetic field, as discussed below. The term $J_0^{xy}$ introduces quantum fluctuations, which can substantially modify the model's ground states (see Sec.~S1 in the SM~\cite{SM}). However, the system can still exhibit SL behavior in the regime $J_0^{xy}< T< J_0^{zz}$. 

\begin{figure*}[t]
\includegraphics[width = \textwidth]{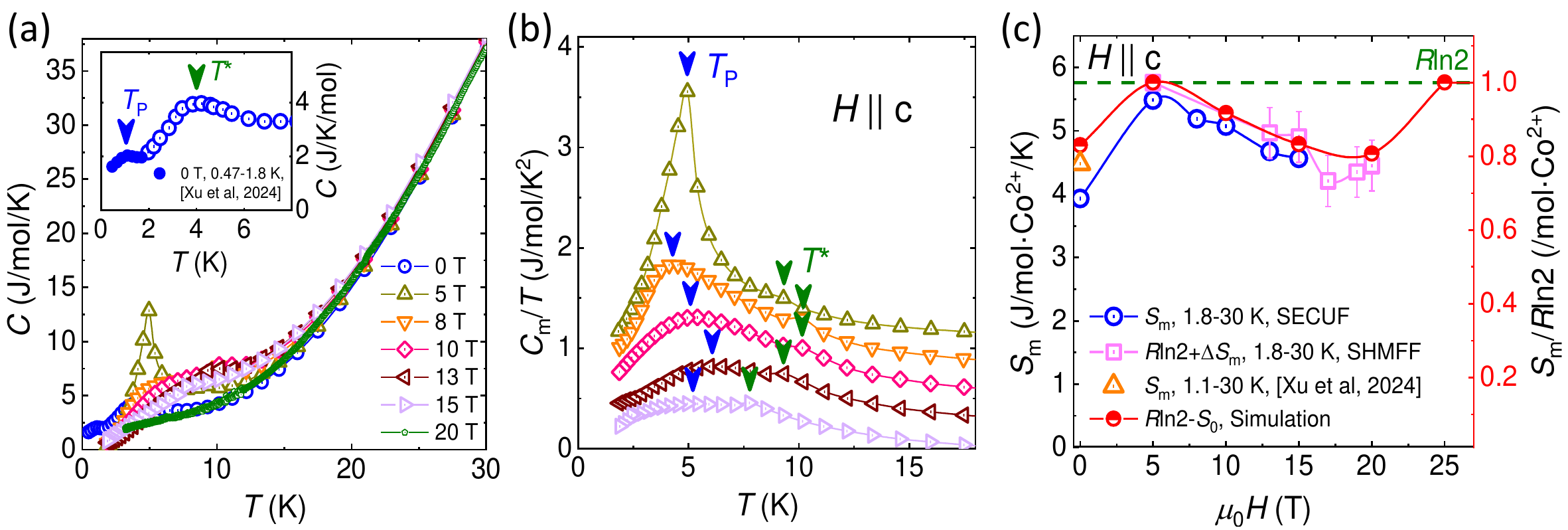}
\caption{\label{cv}{\bf Specific heat and entropy of Rb$_2$Co$_2$(SeO$_3$)$_3$.}
(a) Total specific heat $C$ measured in longitudinal fields. Inset: enlarged view of $C$ at 0~T (Data below 1.8~K are adapted from Ref.~\cite{Xu_2024_CM}). 
(b) Enlarged view of the low-temperature magnetic heat capacity $C_{\rm m}/T$ after subtraction of the phonon contribution (see Sec.~S4 in the SM~\cite{SM}). Vertical offsets are applied for visual clarity. $T^*$ and $T_{\rm P}$ indicate two maxima or kinked features. $T^*$ marks the onset of enhanced spin fluctuations on approaching the plateau phase, and $T_{\rm P}$ marks the temperature at which the plateau phase is reached thermodynamically. 
(c) Magnetic entropy $S_{\rm m}$ obtained by integrating $C_{\rm m}/T$ with data taken in different magnets. The entropy change above 30~K is nearly field-independent (Fig.~S6 in the SM~\cite{SM}).
Data acquired at SHMFF are plotted with the 5~T entropy subtracted (Sec.~S4 in the SM~\cite{SM}). The zero-field magnetic entropy adapted from Ref.~\citenum{Xu_2024_CM} is included for comparison. The half-filled symbols represent the entropy calculated within a classical spin-liquid scenario, after subtracting the residual entropy arising from the macroscopic degeneracy of the unpolarized dimers.
}
\end{figure*}

Using a specific set of parameters, we identify distinct types of ground states with increasing field, corresponding to the four plateau phases mentioned earlier (see Sec.~S1 in the SM~\cite{SM} for details). The ground-state pattern of the 1/3-plateau is illustrated in Fig.~\ref{states}b: one-third of the dimers are occupied by the two-up spin configuration, while the remaining dimers form a honeycomb lattice with antiferromagnetically aligned spins. It follows the UUD pattern in each layer. As the field increases further, the density of antiferromagnetic dimers is diluted, and this effectively reduces the antiferromagnetic correlations of the system, giving rise to magnetically disordered states.  Interestingly, the 2/3-plateau phase adopts a configuration where each antiferromagnetic dimer is surrounded by 12 polarized spins (see Fig.~\ref{states}b). The one-up-one-down rule of each dimer then yields a 2/3-SL with a macroscopic ground-state degeneracy of $2^{N_{\rm d}}$, where $N_{\rm d}$ is the total number of antiferromagnetic dimers. 
We note that the 2/3-SL state is stabilized via a lattice-symmetry breaking effect, evident from the honeycomb lattice formed by the polarized dimers in Fig.~\ref{states}b.

Inclusion of the inter-bilayer $J_3$ interaction stabilizes the 1/2- and 5/6-plateau phases. The 1/2-plateau phase consists of alternating bilayers of UUD and 2/3-SL patterns, having a hybridized nature of antiferromagnetic order and SL. Similarly, one can construct the 5/6-plateau pattern with alternating bilayers of 2/3-SL and the FP configuration. But each antiferromagnetic dimer can hop between neighboring bilayers without costing energy. This yields additional ground-state degeneracy and makes the 5/6-plateau a SL without lattice symmetry breaking (see Sec.~S1 in the SM~\cite{SM}). These ground states from above theoretical analysis are further corroborated by our experimental results on Rb$_2$Co$_2$(SeO$_3$)$_3$ presented below.

In the SL states discussed above, the doubly-degenerate dimer contributes zero to the total magnetization but $k_{\rm B}\ln 2$ to the residual entropy. As a result, the magnetization $M$ and residual entropy in a SL state can be expressed as $M/M_{\rm S}= 1-2N_{\rm d}/N$, and $S_0= N_{\rm d} R\ln2=\frac{1}{2}(1-M/M_{\rm S})R\ln2$, respectively, where $N$ is the total number of lattice sites and $M_{\rm S}$ is the saturation magnetization.  This yields 
$S_{\rm 0}= 1/6R\ln2$ for the 2/3-SL state, and $S_0= (1/12)R\ln2$ for the 1/2-SL (where the 2/3-SL occupies 50\% of the volume). 
For the 5/6-SL, the inter-bilayer hopping of dimers contributes an additional entropy of about $0.11R\ln2$ 
(see Sec.~S1 in the SM~\cite{SM}), which leads to $S_0\approx0.198R\ln2$. The thermal entropy $R\ln 2-S_0$ with the field are plotted in Fig.~\ref{cv}c.
\\

\noindent
{\bf Specific heat}
\\
\noindent
Specific heat $C$ was measured using different magnets in fields of up to 20~T (see Methods and Sec.~S4 of the SM~\cite{SM}). Representative data at selected fields are shown in Fig.~\ref{cv}a. After subtracting the phonon contribution (Sec.~S4 of the SM~\cite{SM}), the magnetic specific heat $C_{\rm m}$ was obtained and is plotted as $C_{\rm m}/T$ in Fig.~\ref{cv}b for fields from 0 to 15~T.

\begin{figure*}[t]
\includegraphics[width=\textwidth]{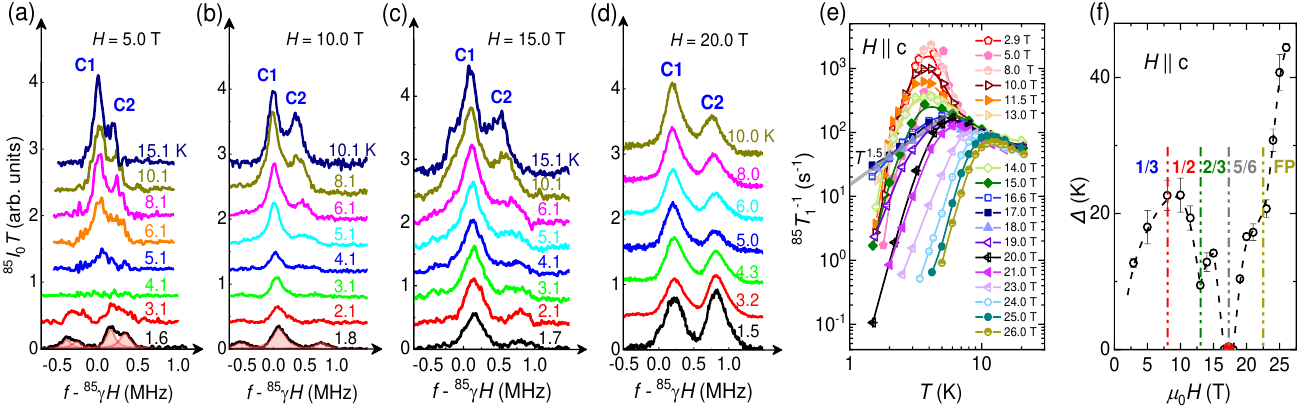}
\caption{\label{nmr}{\bf NMR spectra and spin-lattice relaxation rate.}
(a-d) Typical $^{85}$Rb NMR spectra under the four plateau fields. Spectra at different temperatures are offset vertically for clarity. Peaks C1 and C2 arise from $^{85}$Rb nuclei with two distinct occupancy configurations of neighboring Se(II) sites, respectively (see Fig.~\ref{pd}a). In a and a, four- and three-component Lorentzian fits are applied to the spectra at the lowest temperatures measured at 5~T and 10~T, respectively. The relative spectral weight shows ratios of 1:1:2:2 and 1:4:1, with each fitted component displayed as a red line.
(e) $1/^{85}T_1$ as a function of temperature for increasing magnetic fields. The thick gray line serves as a guide to the eye for $1/T_1 \sim T^{1.5}$ at fields close to 17~T. Enhanced fluctuations are suggested near the transition between the 2/3 and 5/6 plateaus (around 17.3~T) at low temperatures. $1/T_1$ follows a power-law dependence $1/T_1 \sim T^{1.5}$ between 16.6~T and 18~T, indicating the presence of gapless excitations. 
(f) Spin gap $\Delta$ extracted from fitting a thermal activation function to the low-temperature $1/T_1$ data. Vertical dotted lines mark the transition fields between the plateau phases. 
}
\end{figure*}

At 5~T, a sharp peak is observed at approximately 5.0(1)~K, marked as $T_{\rm P}$, which provides clear evidence for a magnetic transition into the UUD phase. Above 5~T, the sharp anomaly at $T_{\rm P}$ is rapidly suppressed and evolves into a broad peak, despite the presence of well-defined magnetization plateaus at 1.8~K (Fig.~\ref{pd}c). Given its continuous evolution with field, $T_{\rm P}$ can be identified as the onset temperature of the plateau phase. Another weak kinked feature in the specific heat, labeled $T^*$, is observed above 7.5~K. Given its similar field dependence to $T_{\rm P}$, it is interpreted as a crossover temperature into the plateau phase.

The magnetic entropy $S_{\rm m}$, calculated from data obtained at SECUF, and the entropy difference $\Delta S_{\rm m}$ (taking the 5~T data as a reference) derived from SHMFF measurements, are plotted in Fig.~\ref{cv}c. At 5~T, $S_{\rm m}$ reaches a value about 5\% lower than $R\ln 2$, likely due to the limited temperature range over which $C_{\rm m}/T$ was integrated. As the field increases, $S_{\rm m}$ is further reduced relative to the 5~T entropy: by about 7\% at 10~T, 15\% at 15~T, and 18\% at 20~T. 
This entropy reduction is inconsistent with an ordered ground state but is instead a hallmark of SLs described above: As shown in Fig.~\ref{cv}c, the field dependence of the measured thermal entropy parallels the calculated one with a difference less than 5\%, which is likely caused by the limited integration temperature range.
\\

\noindent
{\bf NMR spectra}\\
\noindent
Below, we further confirm the existence of SL behavior using NMR data. We performed $^{85}$Rb NMR measurements, with spectra for the four plateau phases presented in Fig.~\ref{nmr}a-d. As shown in Fig.~\ref{nmr}b, the spectra in the high-temperature paramagnetic (PM) phase consist of two center lines (labeled C1 and C2, respectively) with a spectral weight ratio of 1:1. This splitting arises from the disordered Se(II) sites (being either occupied or empty) neighboring to the Rb atoms (see Fig.~\ref{pd}a)~\cite{Zhong_2020_PRB}.

At 5~T in the 1/3-plateau phase (Fig.~\ref{nmr}a), a strong wipe-out of the NMR signal occurs at 4.1~K, indicating a magnetic phase transition. Below 3.1~K, the NMR line splits into components at both positive and negative frequencies relative to $\gamma H$, directly signaling the onset of static antiferromagnetic order. A fit of the spectral function at the lowest temperature (1.6~K) to a sum of four Lorentzians yields a relative spectral weight of 1:1:2:2 from left to right. This weight distribution is fully consistent with an UUD configuration: both the C1 and C2 lines split, and the hyperfine field in the UUD phase produces a line split with a 1:2 relative spectral weight between negative and positive frequencies.

At 15~T in the 2/3-plateau phase (Fig.~\ref{nmr}c) and at 20~T in the 5/6-plateau phase (Fig.~\ref{nmr}d), no significant loss of spectral intensity is observed upon cooling, indicating the absence of a magnetic phase transition. 
Moreover, at temperatures below 2~K$\text{\textemdash}$well within the plateau phases$\text{\textemdash}$no resonance peaks appear at negative frequencies, confirming the complete absence of AFM order.

At 10~T in the 1/2-plateau phase, a magnetic phase transition is again resolved, evidenced by a partial loss of spectral weight at about 4.1~K (Fig.~\ref{nmr}b). On further cooling, the spectrum evolves into three peaks located at negative, near-zero, and positive frequencies, with an overall spectral-weight ratio of $n_1:n_2:n_3\approx 1:4:1$. 
A pure UUD configuration would produce two peaks with a weight ratio $n_1$:2$n_1$ between negative and positive frequencies. The volume ratio between the UUD and the 2/3-SL phases is therefore estimated to be $3n_1:(n_2+n_3-2n_1)\approx 1:1$, which is in excellent agreement with the proposed scenario of alternating UUD and 2/3-SL bilayers and further supports the results of entropy analysis in Fig.~\ref{cv}c.
\\

\noindent
{\bf Spin-lattice relaxation rate}
\\
\noindent
The spin-lattice relaxation rate, $1/T_1$, was measured to probe the low-energy spin fluctuations. Figure~\ref{nmr}e shows $1/^{85}T_1$ measured on the center NMR line in fields up to 26~T. At low fields (2.9–8~T), $1/T_1$ exhibits a pronounced sharp peak, marking the onset of long-range magnetic order. At 15~T and above, the peak in $1/T_1$ becomes strongly broadened. This suppression of low-energy spin fluctuations is consistent with the formation of a spin-liquid state rather than conventional magnetic ordering. A contour plot of $1/^{85}T_1$ is overlaid on the phase diagram, as shown in Fig~\ref{pd}d.

Enhanced spin fluctuations are already visible at high temperature, evolving along the green dotted line towards the 1/3- and 2/3-plateau phases, whereas the 1/2-SL phase emerges only at low temperatures. Below the temperature of the $1/T_1$ peak, the relaxation rate decreases exponentially within all plateau phases, in agreement with the gapped excitations in plateau phases~\cite{Nishimoto_NC_2013,Xu_PRB_2025,Kamiya_NC_2018,Suetsugu_PRL_2024,Zhitomirsky_PTPS_2005}. The spin gap $\Delta$ at each field was extracted by fitting $1/T_1{\sim}e^{-\Delta/K_{\rm B}T}$ (see details in Sec.~S8 of the SM~\cite{SM}) and is plotted in Fig.~\ref{nmr}f. The resulting gaps display several dome-like features, with maxima located at the 1/2-, 2/3-, and 5/6-plateaus. The gap remains large at the 1/3-1/2 and 1/2-2/3 plateau transitions, consistent with numerical simulations and spectral data indicating that the 1/2-SL phase comprises alternating bilayers of UUD order and 2/3-SL. On the other hand, a gapless behavior is observed at the 2/3-5/6 plateau transition field, which implies the existence of a quantum critical point and a candidate quantum spin liquid for the 5/6-plateau phase as discussed below, which warrants further investigation at lower temperatures.
\\

\noindent
{\large \bf Discussions}
\\
\noindent
Our combined experimental and theoretical results demonstrate that Rb$_2$Co$_2$(SeO$_3$)$_3$ hosts a cascade of SLs in the 1/2-, 2/3- and 5/6-magnetization plateau phases at high magnetic fields. These SLs are governed by the one-up-one-down ice rule of a dimer, analogous to the two-in-two-out one in pyrochlore spin ices~\cite{Pauling_JACS_1935,Ramirez_Nature_1999}. The dilution of dimers by the applied field, combined with lattice-symmetry breaking at fractional magnetizations, stabilizes the SLs as a series of incompressible magnetization plateaus.

Interestingly, even at zero field, the entropy obtained by integrating $C_{\rm m}/T$ from $T_{\rm P}$ to 30~K shows a deficit of about 22\%. This suggests a three-sublattice state where 
2/3 of dimers are antiferromagnetically ordered and the remaining 1/3 are Ising disordered~\cite{Chen_PRL_2024} with a residual entropy $S_0 = 1/6R\ln 2$, accounting for the major part of the observed entropy loss. We expect the system is eventually ordered below $T_{\rm P}$ (Fig.~\ref{pd}d) when quantum fluctuations are included, but whether it orders to the hybridized singlet and antiferromagnetic state~\cite{Chen_PRL_2024}, or the Y-type supersolid phase below a Berezinskii-Kosterlitz-Thouless (BKT) phase~\cite{Zhong_PRM_2020,Zhu_npj_2025,Fu_ArXiv_2025} as sketched in Fig.~\ref{pd}, deserves further exploration.

When quantum fluctuations are turned on by further lowering the temperature, the antiferromagnetic Ising dimer condenses to a spin singlet. Given the lattice symmetry breaking in the 1/2- and 2/3-plateaus, we expect the corresponding SLs eventually develop valence bond solid orders. The exception is the 5/6-SL: since it does not break lattice symmetry, the ground state could be a QSL by superposition of extensive singlet configurations$\text{\textemdash}$an intriguing prospect for future ultra-low-temperature experiments. Despite several known cases such as the 1/9 magnetization plateau on kagome lattice antiferromagnet~\cite{Suetsugu_PRL_2024,jeon_np_2024,Zheng_PRX_2025} and the Kitaev model with an in-plane field~\cite{Sears_PRB_2017,Zheng_PRL_2017,yokoi_sci_2021}, spin liquids are mostly studied at zero magnetic field. Our results on the bilayer TLAFM offer a completely different high-field route to spin liquid states, in either classical or quantum regime. Although the dimer ice rule gives rise to a SL behavior with extensive residual entropy, a QSL may emerge out with the aid of frustrated interactions~\cite{Liu_PRB_2016,Hayami_crystals_2025,smith_PRL_2025} via the mechanism discussed above.\\

\noindent
{\small \bf Conflict of interest}
\\
\noindent
The authors declare that they have no conflict interest.
\\

\noindent
{\small \bf Acknowledgments}

\noindent
This work is supported by the National Key Research and Development Program of China (Grant No. 2023YFA1406500 and 2025YFA1412100), the National Natural Science Foundation of China (Grant No.~12134020, No.~12374156, and No.~12334008), and the Scientific Research Innovation Capability Support Project for Young Faculty (Grant No.~ZYGXQNJSKYCXNLZCXM-M26). Part of the specific heat and High-field NMR were supported by the Synergetic Extreme Condition User Facility (SECUF, https://cstr.cn/31123.02.SECUF). We thank the WM2 of the Steady High Magnetic Field Facility (SHMFF, https://cstr.cn/31125.02.SHMFF.WM2) for assistance with the magnetization measurements, and the WM5 of the SHMFF (https://cstr.cn/31125.02.SHMFF.WM5) for assistance with the specific heat measurements. We also acknowledge the support from the Steady High Magnetic Field Facility Instrument and Equipment Renovation.
\\

\noindent
{\small \bf Author contributions.}

\noindent
The project was conceived by W.Y. and Y.C. The samples used in this study were synthesized by X.G. NMR measurements were carried out by X.X., Z.W., Y.C., X.B., R.B., X.S., K.D., J.L., S.L., R.Z., and J.Y. High-field magnetization measurements were carried out by J.W., X.G., C.X., and L.L. Specific heat were carried by J.L., H.R. K.S., and Z.W. Monte Carlo simulations were performed by Y.W., J.D., and R.Y. The manuscript was written by W.Y., X.X., Y.C., J.W. and R.Y., with input from all authors. All authors discussed the results and contributed to the final version of the manuscript.\\

\noindent
{\small \bf Data availability}
\\
\noindent
The data that support the findings of this study are available from the corresponding author upon request. 
\\

\noindent
{\small \bf Code availability}
\\
\noindent
The code is available from the corresponding author upon reasonable request.
\\

\bibliography{rbco223}

\end{document}